\documentclass[]{emulateapj}

\newcommand{\BP}{Ballesteros-Paredes}

\newcommand{\cs}{c_{\rm s}}
\newcommand{\Dv}{\Delta_{\rm v}}

\newcommand{\kms}{{\rm ~km~s}^{-1}}

\newcommand{\Mdense}{M_{\rm dense}}
\newcommand{\Mvir}{M_{\rm vir}}
\newcommand{\Ms}{M_{\rm s}}
\newcommand{\Mstar}{{M_*}}
\newcommand{\Msun}{M_\odot}
\newcommand{\nsf}{n_{\rm SF}}
\newcommand{\pcc}{{\rm ~cm}^{-3}}

\newcommand{\tff}{t_{\rm ff}}

\newcommand{\Tsf}{T_{\rm SF}}

\newcommand{\vrms}{v_{\rm rms}}
\newcommand{\VS}{V\'azquez-Semadeni}

\newcommand{\msun}{\mbox{$M_\odot$}}

\begin{document}

\title{Molecular Cloud Evolution III. Accretion {\it vs.} stellar feedback} 
\shorttitle{Molecular Cloud Evolution III. Accretion {\it vs.} Stellar Feedback}

\author{Enrique \VS, Pedro Col\'in, Gilberto C.\ G\'omez, and
Alan W. Watson\altaffilmark{1}}
\affil{Centro de Radioastronom\'\i a y Astrof\'\i sica,
Universidad Nacional Aut\'onoma de M\'exico, Campus Morelia,
Apdo. Postal 3-72, Morelia, 58089, M\'exico}
\email{e.vazquez@crya.unam.mx, p.colin@crya.unam.mx, g.gomez@crya.unam.mx,
alan@astro.unam.mx}

\altaffiltext{1}{Also: Instituto de Astronom\'\i a,
Universidad Nacional Aut\'onoma de M\'exico, 
A.P. 70-264, 04510, M\'exico, D.F., M\'exico}

\shortauthors{\VS\ et al.}



\label{firstpage}

\begin{abstract}

We numerically investigate the effect of feedback from the ionizing
radiation heating from massive stars on the evolution of giant molecular
clouds (GMCs) and their star formation efficiency (SFE), which we treat
as a time-dependent quantity. We consider the GMCs' evolution from their
formation by colliding warm neutral medium (WNM) streams up to advanced
star-forming stages. We find, in agreement with our previous studies,
that the star-forming regions (``clouds'') within the GMCs are
invariably formed by gravitational contraction, so that their internal
non-thermal motions must contain a significant component of global
convergence, which does not oppose gravity. After an initial period of
contraction, the collapsing clouds begin forming stars, whose feedback
evaporates part of the clouds' mass, opposing the continuing accretion
from the infalling gas. The competition of accretion against dense gas
consumption by star formation (SF) and evaporation by the feedback,
regulates the clouds' mass and energy balance, as well as their SFE. We
find that, in the presence of feedback, the clouds attain levels of the
SFE that are consistent {\it at all times} with observational
determinations for regions of comparable SF rates (SFRs).  However, we
observe that the dense gas mass is {\it larger} in general in the
presence of feedback, and that the total (dense gas + stars) is nearly
insensitive to the presence of feedback, suggesting that the total mass
is determined by the accretion, while the feedback inhibits mainly the
conversion of dense gas to stars, because it acts directly to reheat and
disperse the gas that is directly on its way to forming stars.

We find that the factor by which the SFE is reduced upon the inclusion
of feedback is a decreasing function of the cloud's mass, for clouds of
size $\sim 10$ pc. This naturally explains the larger observed SFEs of
massive-star forming regions. We also find that the clouds may attain a
pseudo-virialized state, with a value of the virial mass very similar to
the actual cloud mass. However, this state differs from true
virialization in that the clouds are the center of a large-scale
collapse, continuously accreting mass, rather than being equilibrium
entities. Finally, we also calculate the density probability density
functions of the clouds, finding that they in general exhibit the
bimodal shape characteristic of thermally bistable flows, rather than a
lognormal form, which is characteristic of isothermal flows. This
supports suggestions that low density, atomic gas pervades
molecular clouds. We conclude that the general state of star-forming
regions is likely to be one of gravitational collapse, although most of
the mass in a GMC may be not participating of the instantaneous
star-forming activity, as recently suggested by \citet{Elm07}; that is,
that SF is a spatially and temporally intermittent phenomenon, with
strong, localized bursts interspersed within much more quiescent gas.

\end{abstract}
 

\section{Introduction} \label{sec:intro}
The evolution of giant molecular clouds (GMCs) is a
key ingredient in our understanding of the star formation process.
In particular, the low observed star formation efficiency (SFE) at the
scale of whole GMCs \citep[$\sim $ 2\%;][]{Myers_etal86} remains a topic
of strong debate, with there being two main competing scenarios that
attempt to explain it. These scenarios refer essentially to the effect
of the stellar feedback (mainly from massive stars) on the star-forming
clouds. One is the scenario that the stars quickly
disrupt their parent clouds by dispersal and/or photoionization, before
the gaseous mass of the cloud is completely converted to stars
\citep[e.g.][]{Whit79, Elm83, FST94, WM97, HBB01}. In this case, the SFR
of active star-forming sites may be large for brief periods, and then
halted by the very stars that have just been formed.

In the other scenario, the role of stellar
feedback is to drive turbulent motions within the GMC, which
oppose its self-gravity, allowing it to remain in near hydrostatic
equilibrium for times significantly longer than its free-fall time
($\tff$) \citep{KM05, KMM06, LN06}. In this case, the low efficiency of
star formation would be due to the dual role of supersonic turbulence in
self-gravitating clouds, of opposing global collapse of the cloud while
promoting local collapse of turbulent density enhancements, which
involve small fractions of the total cloud mass \citep[][ see also the
reviews by Mac Low \& Klessen 2004; \BP\ et al. 2007]{KHM00, VBK03}. 

Another controversy, related to the control of the SFE, refers to the
nature of the motions originating the linewidths observed in GMCs and
their substructure. The latter were initially proposed to correspond to
gravitational contraction by \citet{GK74}, but this suggestion was
quickly deemed untenable by \citet{ZP74}, who noted that it would imply
total Galactic SFRs of the order of the total molecular gas mass in the
Galaxy ($\sim 10^9 \Msun$) divided by the typical free-fall time for a
GMC ($\sim 4$ Myr), or $\sim 250 \Msun$ yr$^{-1}$, an estimate roughly
two orders of magnitude larger than the observed Galactic SFR.
\citet{ZE74} then suggested that the observed linewidths could
correspond instead to random, small-scale\footnote{\citet{ZE74} referred
to these motions as ``local'', and explicitly discarded large-scale
coherent motions such as gravitational contraction.}  turbulent motions,
a notion that has prevailed until the present. However, a number of
workers have recently advocated a return to the gravitational
contraction picture, noting that various observational properties of
clouds and clumps can be well matched by models dominated by
gravitational contraction \citep[e.g.,][]{HaBu07, PHA07,
VS_etal09}. Moreover, the notion of completely random, small-scale
turbulent motions appears difficult to reconcile with the recent
realization that the principal component of the velocity differences
within clouds and clumps at all scales appears to be ``dipolar'',
indicative of coherent motions at the scale of the whole cloud or clump
\citep{HeBr07, BHM09}. In general, several studies comparing simulations and
observations have concluded that the motions in molecular clouds are
consistent with large-, rather than small-scale driving \citep{OM02,
Brunt03, Padoan_etal09}.

If a return to the collapsing scenario is to be considered, it is
necessary to somehow avoid the \citet{ZP74} criticism of it.  This is
actually not so difficult because that criticism neglects the internal
structure of the GMCs. Recent numerical studies of cloud formation by
converging streams of warm neutral gas in the interstellar medium show
that the clouds are born turbulent, due to one or more of the thermal,
thin-shell, and Kelvin-Helmholz instabilities \citep{HP99, KI00, KI02,
AH05, Heitsch_etal05, VS_etal06}. The turbulence is subsonic with
respect to the warm gas, but supersonic with respect to the cold phase,
implying large density fluctuations in the latter.
In fact, it has recently been proposed that molecular clouds may
actually contain a warmer, atomic substrate in which colder molecular
clumps are embedded \citep{HI06}. In either case, the molecular cloud
contains large, {\it nonlinear} density enhancements in which the local
free-fall time is significantly shorter than the cloud's average. Thus,
once the global collapse begins, the local clumps may complete their
collapses earlier than the bulk of the cloud. They can thus begin
forming stars that can begin their feedback action on the GMC before it
completes the bulk collapse. 

In this paper, we present numerical simulations aimed at investigating
this scenario, in which we use the same cloud-formation setup of
previous papers \citep{VS_etal07, Banerjee_etal09}, but including a
prescription for stellar feedback mimicking ionization heating from
massive stars. With this tool, we investigate the effect of the feedback
on the global SFE of the evolving GMC, as well as the nature of the
motions in the cloud, in a first effort to shed light on these issues.
As we shall see, it turns out that the physical conditions in the clouds
differ significantly from the ``normal'' picture, since accretion of gas
from the warm diffuse medium is an integral part of the clouds' dynamics
and evolution, and thus the clouds cannot be considered as isolated.

The plan of the paper is as follows. In \S \ref{sec:model} we describe
the numerical code, and the implementation of our star formation and
stellar feedback prescriptions. In \S \ref{sec:simulations} we describe
the simulations, and in \S \ref{sec:results} we describe the results
concerning the control of the SFE by stellar feedback and the nature of
the ``clouds'' themselves. Finally, in \S \ref{sec:conclusions} we
present a summary and some conclusions.

\section{The numerical model} \label{sec:model}

\subsection{Heating and cooling} \label{sec:code}

The numerical simulations used in this work were performed using 
the hydrodynamics + N-body Adaptive Refinement Tree code ART 
\citep[]{KKK97,Kravtsov03}. Among the physical 
processes implemented in ART, relevant for our physical problem, are the
radiative heating and cooling of the gas, its 
 conversion into stars, ionization-like heating from stellar feedback, and
self-gravity, from both the stars and the gas.

We use heating ($\Gamma$) and cooling ($\Lambda$) functions of the form  

\begin{eqnarray} 
\Gamma &=& 2.0 \times 10^{-26} \hbox{ erg s}^{-1}\label{eq:heating}\\
\frac{\Lambda(T)}{\Gamma} &=& 10^7 \exp\left(\frac{-1.184 \times
10^5}{T+1000}\right) \nonumber \\
&+& 1.4 \times 10^{-2} \sqrt{T} \exp\left(\frac{-92}{T}\right)
~{\rm cm}^3.\label{eq:cooling}
\end{eqnarray}
These functions are fits to the various heating and
cooling  processes considered by
\citet{KI00}, as given by equation (4) of \citet{KI02}. As noted in
\citet[][hereafter Paper I]{VS_etal07}, eq.\ (4) in \citet{KI02}
contains two typographical 
errors. The form used here incorporates the necesary corrections,
kindly provided by H.\ Koyama (2007, private communication).
With these heating and cooling laws, the gas is thermally unstable in
the density range $1 \la n \la 10 \pcc$ (cf. Paper I). 

\subsection{Star formation and stellar feedback prescriptions}
\label{sec:SF_model}  

In ART, star formation is modeled as taking place in the coldest
and densest regions, defined by $T < \Tsf$ and $n >
\nsf$, where $T$ and $n$ are the local temperature and number density of
the gas, respectively, and $\nsf$ and $\Tsf$ are respectively a density
and a temperature threshold. We set $\Tsf = 9000$ K, which is easily
satisfied by all cells with density $\nsf$, so in practice our
SF condition depends on density only. 

A stellar particle of mass $m_*$ is placed in a grid cell where these
conditions are simultaneously satisfied, and this mass is removed from
the gas mass in the cell. Thereafter, the particle is treated as
non-collisional, and follows N-body dynamics. No other criteria are
imposed. In each gas cell that satisfies the above criteria a stellar
particle is formed with a mass equal to 50\% of the gas mass contained
in the cell. Since the stellar particles are more massive than a single
star, each stellar particle should be considered as a small cluster,
within which the individual stellar masses are distributed according to
some stellar initial mass function (IMF).

Stellar particles inject thermal energy at a rate  $\dot E$ erg
Myr$^{-1}$ per star with mass greater than $8\ \msun$ contained in the
stellar particle. We assume a \citet{MS79} IMF,
implying that each stellar particle of mass $133~\msun$ produces one
8-$\Msun$ star. The energy is deposited in the cell in which the
stellar particle is instantaneously located, over a typical OB stellar
lifetime, which we assume to be 10 Myr.

It is important to note that, although initially we experimented with
realistic values of $\dot E$ based on the Lyman continuum
fluxes of stars with masses between 10 and 20 $\Msun$
\citep[e.g.,][]{Diaz_etal98}, we found that, because all the energy is
deposited in a single cell, and the neighboring cells are heated
exclusively by conduction, rather than by radiative heating, the
resulting HII regions were not so realistic. Thus, we opted instead for
taking $\dot E$ as a free parameter, and adjusting it until we obtained
realistic HII regions, with temperatures $ \sim 10^4$ K, diameters of a
few parsecs, and expansion velocities of a few tens of$\kms$.

Note also that we resort to the common strategy of turning off the
cooling in the cell where a stellar particle is located, so that the
cell can reach realistically high temperatures. Otherwise, the cooling
can dissipate most of the thermal energy deposited in very dense cells.
In the real ISM this does not occur because the stellar heating is
applied through ionization, so that the temperature reached in the
star's immediate environment is independent of the medium's local
density. Instead, in the simulations, the cooling depends on the
density, and the temperature resulting from the balance between the
stellar heating and the cooling does depend on the density. This problem
is avoided by turning off the cooling in the cell where the stellar
particle is located. Note that this contradicts claims that the need to
turn off the cooling can be alleviated simply by increasing the
resolution \citep[e.g.,][]{CK09}. We argue that this problem can only be
alleviated by performing radiation-hydrodynamics simulations. In their
absence, we consider that turning off the cooling is actually a {\it
better} model of the effect of stellar feedback, because it allows
mimicking the fact that the gas temperature in the HII regions is
independent of the local density.

Finally, in order to further constrain the physical conditions in the
HII regions, we also impose a ``ceiling'' to the temperature in the cell
containing the stellar particle because otherwise, with the cooling
off, the temperature in the cell might diverge. We set this ``ceiling''
to $10^6$ K.

\begin{figure*}
\plotone{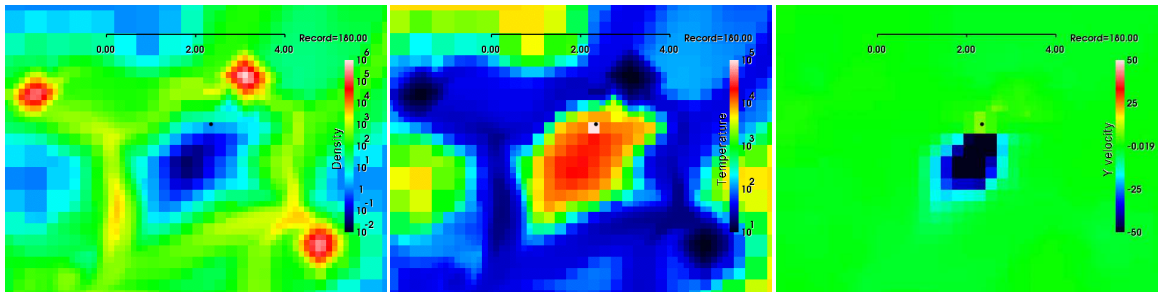}
\caption{Cross sections of the density ({\it left panel}), temperature
({\it middle panel}) and $y$-speed ({\it right panel}), shown on the $x$-$z$
plane, of a typical isolated HII region. The scale bar near the top
indicates length in parsecs.}
\label{fig:HII_region}
\end{figure*}

Although this procedure is mostly one of trial-and-error, we consider it
to be the most adequate one for our purposes, since it is the HII
regions that drive the turbulent motions in the dense gas in our
simulations, and so it is them that must have realistic properties, even
at the expense of a somewhat ad-hoc SF prescription. We show a typical
HII region in our simulation in Fig.\ \ref{fig:HII_region}.

\subsection{Refinement} \label{sec:refinement}

The numerical box is initially covered by a grid of $128^3$ (zeroth
level) cells. The mesh is subsequently refined as the matter
distribution evolves. The maximum allowed refinement level was set to
four, so that high-density regions have an effective resolution of
$2048^3$ cells, with a minimum cell size of 0.125 pc. Cells are refined
(halved in linear size) if the gas mass within the cell is greater than
0.32 \msun. That is, the cell is refined by a factor of 2 when the
density increases by a factor of 8, so that, while refinement is active,
the grid cell size $\Delta x$ scales with density $n$ as $\Delta x
\propto n^{-1/3}$. Once the maximum refinement level is reached, no
further refinement is performed, and the cell's mass can reach much
larger values. In particular, a stellar particle is formed when a
fourth-level cell reaches a density $\nsf = 4 \times 10^6 \pcc$, or a
mass of $243.5 \Msun$, again assuming $\mu = 1.27$ (we use this value
because we do not follow the actual chemistry, and thus we assume the
entire box to be filled with atomic hydrogen.) Thus, a stellar particle
typically has a mass $\ga 122 \Msun$. 

Note that, because we use only four levels of refinement, the largest
densities arising in the simulation are by far not sufficiently resolved
according to the ``Jeans criterion'' proposed by \citet{Truelove_etal97}
for adaptive-mesh codes. Specifically, at our stellar-particle formation
threshold density of $4 \times 10^6 \pcc$, and assuming a gas
temperature of $ T \sim 15$ K at that density, we find that the Jeans
length (using the adiabatic sound speed) is $\sim 0.031$ pc, while the
minimum cell size, at 0.125 pc, is roughly 4 times larger. Thus,
according to those authors, one should expect artificial fragmentation
to occur in our simulations. However, we do not consider this to be a
problem because we are not concerned here with the numbers and masses of
the stellar particles formed in the simulation, but simply with the {\it
total} amount of mass that goes into stars.

\section{The simulations} \label{sec:simulations}

We consider four simulations using the same setup as in Paper I, which
represents the evolution of a region of 256 pc per side, initially
filled with warm gas at a uniform density of $n_0 = 1 \pcc$ and a
temperature $T_0 = 5000$ K, implying an adiabatic sound speed $\cs = 7.4
\kms$ (assuming a mean particle mass $\mu =1.27$). The whole numerical
box thus contains $5.25 \times 10^5 \Msun$. In this medium, we set up
two streams moving with the same speed $v_{\rm inf} = 5.9 \kms$
(corresponding to a Mach number of 0.8 with respect to the unperturbed
medium) in opposite senses along the $x$-direction. The streams have a
radius of 32 pc and a length of 112 pc each, so that the total mass in
the two inflows is $2.25 \times 10^4 \Msun$. The flows collide head on
at the box's center (see Fig.\ 1 of Paper I). To the inflow velocity
field we superpose a field of initial low-amplitude turbulent velocity
fluctuations, in order to trigger the instabilities in the compressed
layer. We create this initial velocity fluctuation field with a new
version of the spectral code used in \citet{VPP95} and \citet{PVP95},
modified to run in parallel in shared-memory architectures.  The
simulations are evolved for about 40 Myr.

The collision nonlinearly triggers a transition to the cold phase,
forming a turbulent, cold, dense cloud \citep{AH05, Heitsch_etal05,
Heitsch_etal06, VS_etal06}, consisting of a complex network of sheets,
filaments, and clumps of cold gas embedded in a warm diffuse substrate
\citep{AH05, HI06, HA07}. The largest cold structures may become
gravitationally unstable and begin to collapse. Eventually, they proceed
to forming stars, which then heat their environment, forming expanding
``HII regions'' that tend to disperse the clumps.

In the simulations reported here, we vary only two parameters: the
amplitude of the initial velocity fluctuations and whether the stellar
feedback is on or off. We consider a ``large-amplitude'' (LA) and a
``small-amplitude'' (SA) case for the initial velocity fluctuations, for
which the three-dimensional velocity dispersions are $\vrms \sim 1.7
\kms$ and $\vrms \sim 0.1 \kms$, respectively. We thus employ a mnemonic
nomenclature for the runs using the acronyms LA or SA, followed by F0 or
F1, indicating that feedback is off or on, respectively. 
Table~\ref{tab:run_parameters} summarizes the runs considered in the paper.

\begin{deluxetable}{ccc}
\tablecaption{\sc Run parameters}
\tablehead{\colhead{Run} & \colhead{$\vrms$} & \colhead{Feedback} \\
  name	& [km s$^{-1}$] &	}	\\
\startdata
LAF0 	& 1.7		& off		\\
LAF1 	& 1.7		& on		\\
SAF0 	& 0.1		& off		\\
SAF1 	& 0.1		& on		\\
\enddata
\label{tab:run_parameters}
\end{deluxetable}

%

\section{Results} \label{sec:results}

\subsection{Evolution of the simulations} \label{sec:evolution}

The simulations performed here behave very similarly to previous
simulations with similar setups, performed with other codes. In
particular, our SA runs are very similar to run L256$\Delta v$0.17 in
Paper I and the run presented by \citet{Banerjee_etal09}. The main
feature of these runs is that, because the initial fluctuations are very
mild, the flow collision creates a large, coherent ``pancake'' of cold,
dense gas, which is able to undergo gravitational collapse as a whole.
This results in the formation of a dense, massive, and turbulent region
at the site where the global collapse finally converges, with physical
properties similar to those of high-mass star forming regions
\citep{VS_etal09}. However, a recent study varying the parameters of the
flow collision \citep{Rosas_etal10} shows that the coherence of the
collapse may be lost in the presence of stronger initial velocity
fluctuations, and the SFE is decreased.  In such cases, smaller clouds
appeared to be less strongly gravitationally bound, with the effect of
decreasing the SFE. This feature also happens in our LA simulations, in
which the cloud formed by the initial flow was much more fragmented and
scattered over the simulation volume. As a result, SF also occurs in a
much more scattered manner, and the SFEs are in general smaller in the
LA runs than in their SA counterparts.
\begin{figure}
\plotone{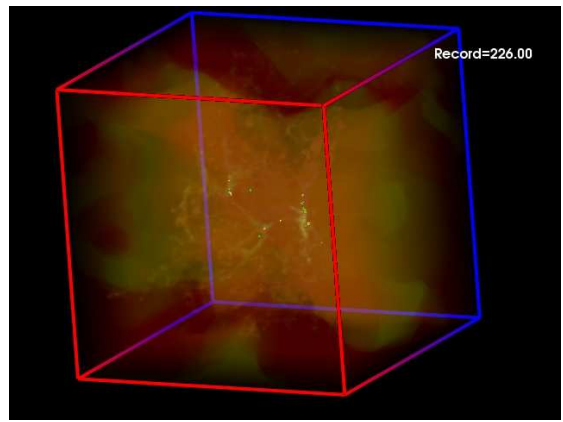}
\caption{View in projection of the whole numerical box for simulation
LAF1 at $t \approx 31.64$ Myr. The box size is 256 pc. The green dots
indicate stellar particles. The light yellow spots are transient dense
cores, highlighted by saturation of the color table.  In the electronic
version, this figure shows an animation of the entire evolution of the
simulation up to $t=40$ Myr. Records are spaced by a time interval
$\Delta t \approx 0.14$ Myr. }
\label{fig:LAF1_whole_proj}
\end{figure}

However, in general a common pattern is followed by all simulations:
the transonic converging flows in the diffuse gas induce a phase transition
to the cold phase of the atomic gas, which is highly prone to
gravitational instability. This can be seen as follows. The thermal
pressure at our initial conditions is 5000 K$\pcc$. From Fig.\ 2 of
Paper I, it can be seen that the thermal balance conditions of the cold
medium at that pressure are $n \sim 130 \pcc$, $T \sim 40$ K. At
these values, the Jeans length and mass are $\sim 7$ pc and $\sim 640
\Msun$, respectively. These sizes and masses are easily achievable by a
large fraction of the cold gas structures, which can then proceed to
gravitational collapse and form stars. Moreover, the ensemble of these
clumps may also be gravitationally unstable as a whole, the likelihood
of this being larger for greater coherence of the large-scale pattern.

Regions of active star formation form in both sets of simulations by the
gravitational merging of pre-existing smaller-scale clumps, which,
altogether, form a larger-scale GMC. Figure \ref{fig:LAF1_whole_proj}
shows a whole-box image of the density field in run LAF1, in
projection. In the electronic version of the paper, this figure shows an
animation of this run from $t=0$ to $t\approx 40$ Myr, illustrating the
entire evolution of the simulation, from the assembly of the cloud, to
its advanced star-forming epochs. In the animation, subsequent
``records'' are separated by time intervals $\Delta t \approx 0.14$ Myr.

\begin{figure}
\plotone{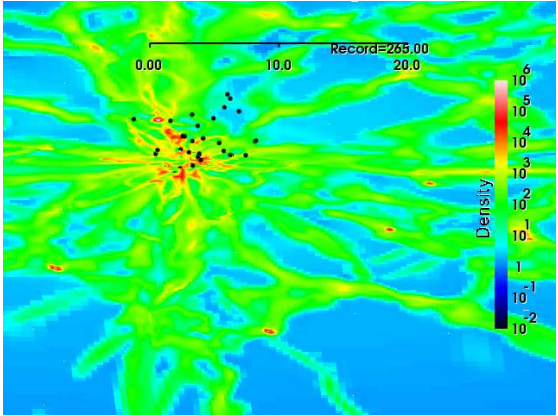}
\caption{Cross-section view through the Central Cloud in run SAF1 at $t
\approx 33$ Myr, at which time it has grown to a mass of $\sim 3 \times
10^4 \Msun$. The plane of the image is shown from an inclined line of
sight for better perspective. The cloud is seen to contain numerous HII
regions mixed with dense regions. In the electronic version, this figure
shows an animation of the build-up of this cloud, illustrating how it
forms by the continued accretion of infalling material from the globally
collapsing GMC.  In the animation, note that many of the infalling
clumps form stellar particles before reaching the center, and are
disrupted by the local stellar feedback. However, once the Central Cloud is
fully assembled, it resists dispersal, and forms stars at a high rate.}
\label{fig:SA_CC_t265_zoom}
\end{figure}

In the SA runs, the largest
star-forming region forms in the center of the simulation, due to the
coherent collapse of the entire sheet-like cloud formed by the
collision. This was the region shown in \citet{VS_etal09} to
exhibit physical conditions typical of actual high-mass star-forming
regions. We refer to this region as ``the Central Cloud''.  Figure
\ref{fig:SA_CC_t265_zoom} shows a view of this region at $t \approx 33$
Myr, a time at which the central cloud has grown to a mass of nearly $3
\times 10^4 \Msun$ in run SAF1 (cf.\ Fig.\
\ref{fig:SA_gas_star_mass_evol}).
In the case of the LA runs, because star formation occurs in a much more
scattered fashion, we study two of the regions exhibiting the strongest
star formation activity, neither of which is located at the center of
the numerical box. These are shown in Fig.\ \ref{fig:LA_t248_zoom}, and
we refer to them as Cloud 1 and Cloud 2. 

\subsection{Effect of stellar feedback on the SFE and on the clouds'
evolution} \label{sec:SFE} 

\begin{figure}
\plottwo{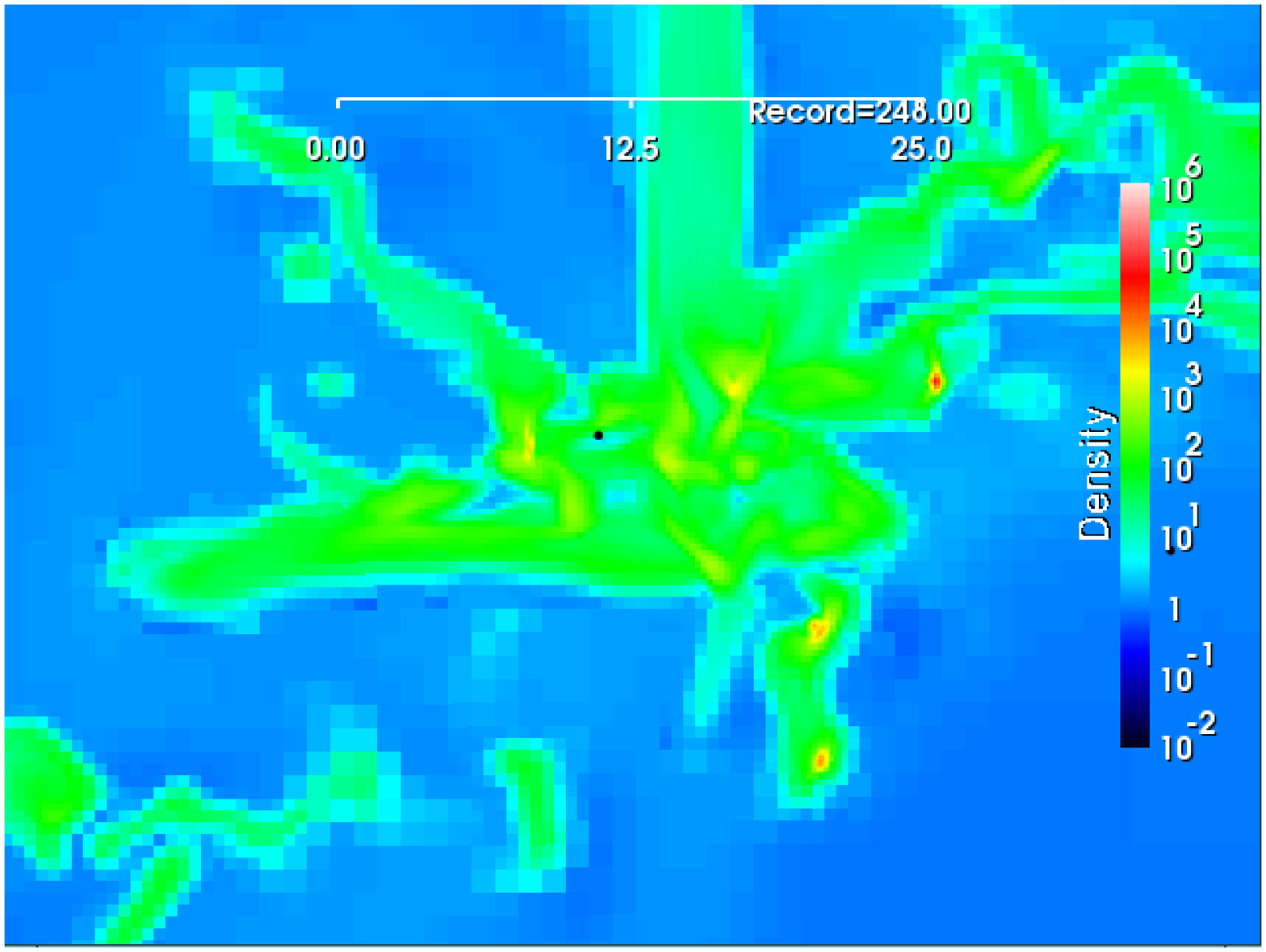}{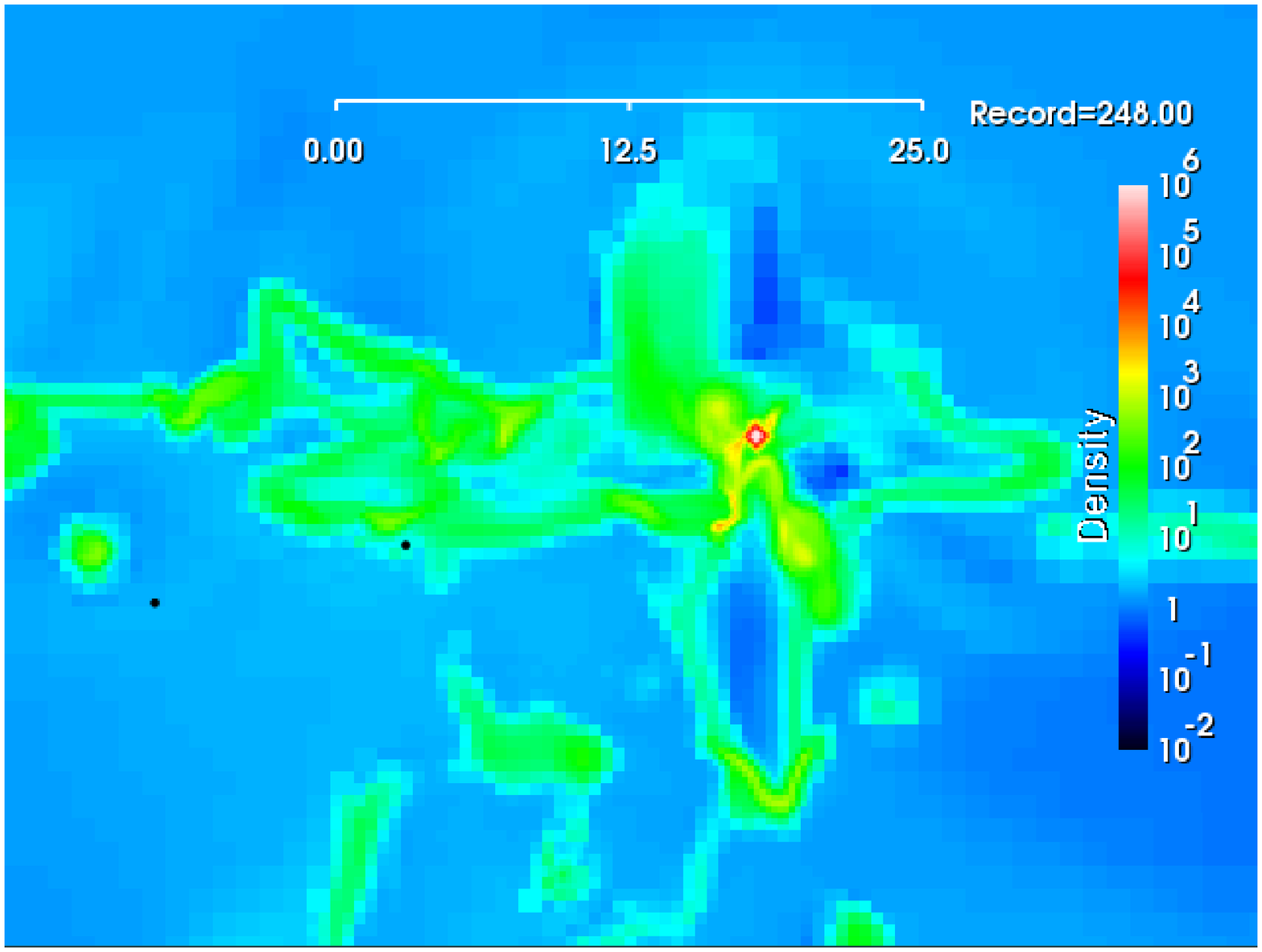}
\caption{Cross-section view through Clouds 1 ({\it top panel}) and 2
({\it bottom panel}) at $t \sim 35$ Myr in
simulation LAF1. The plane of view is located at $x = 100$ pc for Cloud
1 and at $x = 150$ pc for Cloud 2. The dots represent the stellar
particles. In the electronic edition, these figures show animations of
the evolution of both clouds from $t =23$ to $t = 40$ Myr.
}
\label{fig:LA_t248_zoom}
\end{figure}

\begin{figure}
\includegraphics[width=0.5\textwidth]{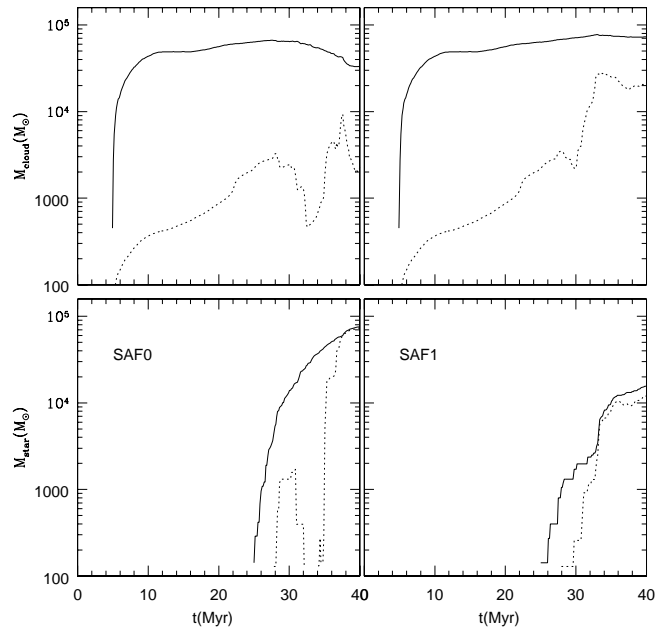}
\caption{Evolution of the dense gas mass and the stellar mass for the
SA simulations, without ({\it left panels}, run SAF0) and with ({\it
right panels}, run SAF1) feedback. The {\it solid} lines refer to the
total masses in the computational box, while the {\it dotted} lines
refer to the masses in a cylinder with length and diameter of 10 parsecs
with its axis along the $x$-direction, and which contains the Central
Cloud.  }
\label{fig:SA_gas_star_mass_evol}
\end{figure}

\begin{figure}
\includegraphics[width=0.5\textwidth]{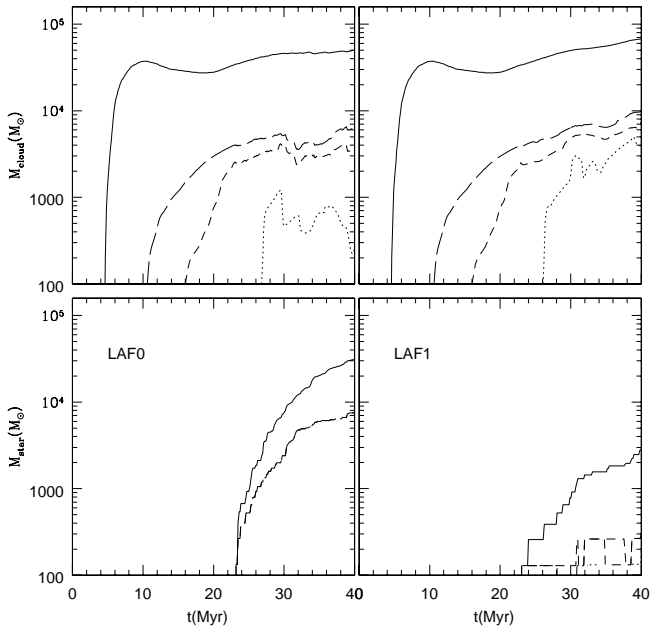}
\caption{Evolution of the dense gas mass and the stellar mass for Cloud
1 in the LA simulations, without ({\it left panels}) and with ({\it
right panels}) feedback. The {\it solid} lines refer to the total masses
in the computational box. The other lines refer to the masses in
cylinders of length and diameter 10 pc ({\it dotted lines}), 20 pc ({\it
short-dashed lines}), and 30 pc ({\it long-dashed lines}).  }
\label{fig:LA_CL1_gas_star_mass_evol}
\end{figure}

\begin{figure}
\includegraphics[width=0.5\textwidth]{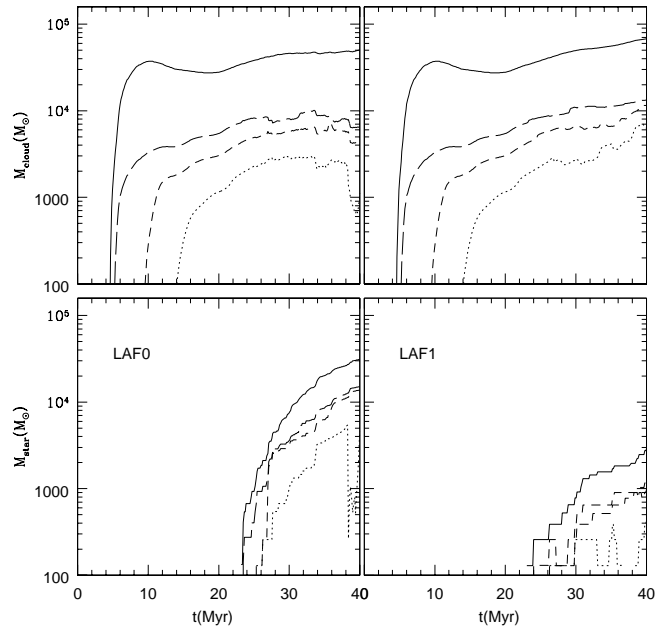}
\caption{Evolution of the dense gas mass and the stellar mass for Cloud
2 in the LA simulations, without ({\it left panels}) and with ({\it
right panels}) feedback. The {\it solid} lines refer to the total masses
in the computational box. The other lines refer to the masses in
cylinders of length and diameter 10 pc ({\it dotted lines}), 20 pc ({\it
short-dashed lines}), and 30 pc ({\it long-dashed lines}).  }
\label{fig:LA_CL2_gas_star_mass_evol}
\end{figure}

Our main interest in this contribution is the effect of the feedback on
the efficiency of the star formation process, and the identification of
the mechanism through which this effect is accomplished.  Figure
\ref{fig:SA_gas_star_mass_evol} shows the evolution of the dense gas
mass and the stellar mass in these simulations, with ({\it right
panels}) and without ({\it left panels}) feedback. The {\it solid} lines
refer to the total masses in the computational box, while the {\it
dotted} lines refer to the masses in the Central Cloud. Figures
\ref{fig:LA_CL1_gas_star_mass_evol} and
\ref{fig:LA_CL2_gas_star_mass_evol} show the corresponding plots for 
Cloud 1 and Cloud 2. Here, the {\it solid lines} represent the masses
for the full simulation box, and the {\it dotted, short-dashed, and
long-dashed lines} represent the masses contained in cylinders of length
and diameter 10, 20, and 30 pc, respectively, enclosing the clouds. We
do this because the clouds have very complicated morphologies, with
filaments that extend out over tens of parsecs and connecting with other
clouds (Fig.\ \ref{fig:LA_t248_zoom}), thus making it virtually
impossible to fully enclose the ``clouds'' in any given cylindrical
volume. 

\begin{figure}
\includegraphics[width=0.6\textwidth]{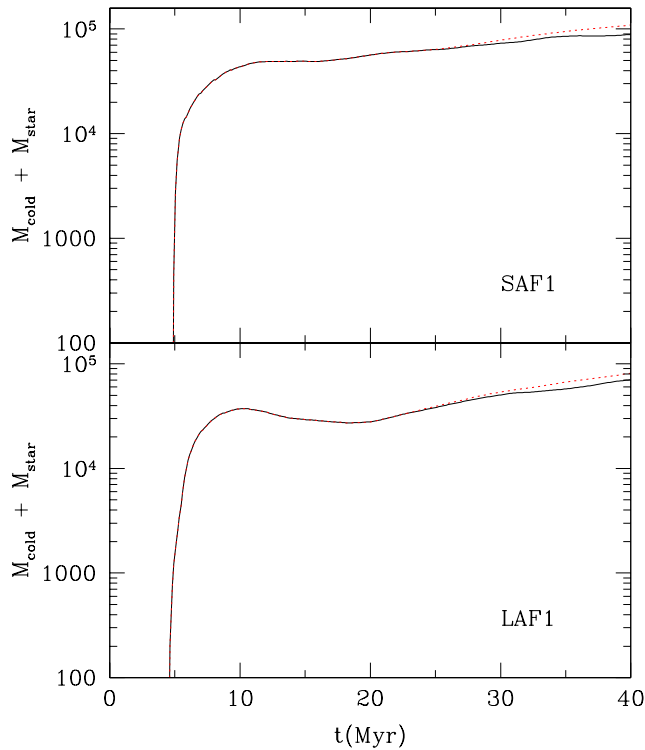}
\caption{Evolution of the total (dense gas + stars) mass for the four
simulations. The SA simulations are shown in the {\it top panel}, while
the LA runs are shown in the {\it bottom panel}. The {\it black, solid}
lines refer to simulations with feedback and the {\it red, dotted} lines
represent runs without it. The colors are shown in the electronic
version only.
}
\label{fig:total_mass_evol}
\end{figure}

\begin{figure}
\includegraphics[width=0.5\textwidth]{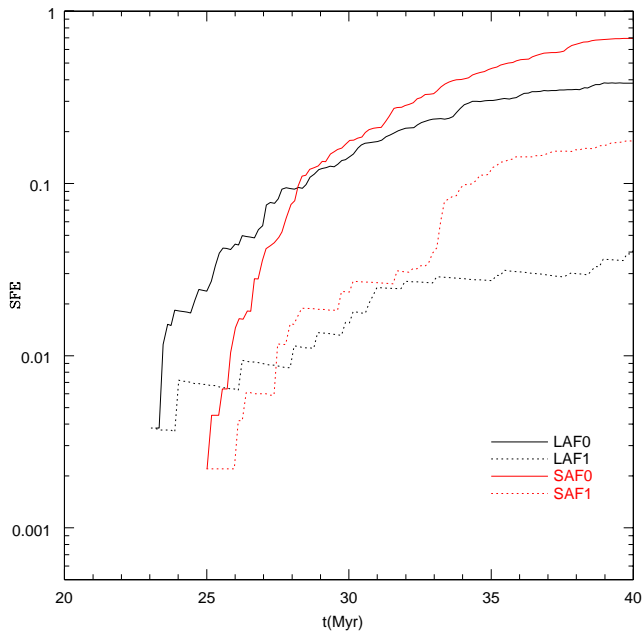}
\caption{Evolution of the instantaneous star formation efficiency (SFE),
as defined in eq.\ (\ref{eq:SFE_def}), in the full
simulation box in the four runs.
}
\label{fig:SFE_global}
\end{figure}

It is seen from Figs.\
\ref{fig:SA_gas_star_mass_evol}-\ref{fig:LA_CL2_gas_star_mass_evol} that
the inclusion of feedback ({\it right panels}) causes the dense gas mass
to be {\it larger} and the stellar mass to be smaller than in the case
without feedback in general, even though the total cloud mass (dense gas
+ stars) in the simulations is nearly the same in both the cases with
and without feedback (Fig.\ \ref{fig:total_mass_evol}). As a
consequence, the instantaneous SFE, defined as
\begin{equation}
\hbox{SFE}(t) = \frac{\Mstar(t)}{\Mdense(t) + \Mstar(t)},
\label{eq:SFE_def}
\end{equation}
where $\Mdense(t)$ is the mass of the dense ($n \ga 50 \pcc$) gas in the
simulation, and $\Mstar(t)$ is the stellar mass, naturally decreases upon
the inclusion of feedback in both sets of simulations (Fig.\
\ref{fig:SFE_global}).  Note that in eq.\ (\ref{eq:SFE_def}) we have
explicitly written out the time dependence of the quantities involved.

\begin{figure*}
\plotone{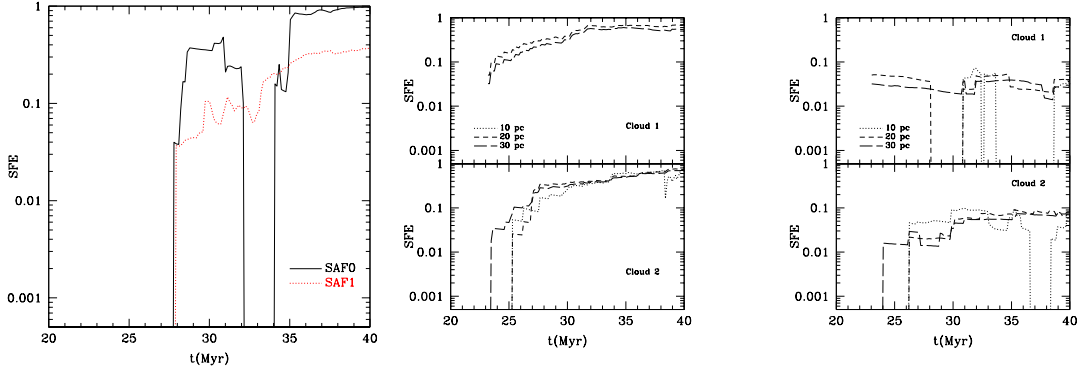}
\caption{Evolution of the instantaneous star formation efficiency (SFE)
in the dense 
clouds in the simulations. The {\it left panel} shows the SFE for the
Central Cloud of the SA runs, with and without feedback. The
measurements refer to a cylindrical box with a diameter and a length of
10 pc. The {\it middle panels} show the SFE for Clouds 1 and 2 in the
LAF0 simulation (without feedback), for three different cylindrical
boxes, of length and diameter indicated in the labels. The gaps in the
curves for the smaller cylindrical boxes correspond to times when the
stellar particles migrate out of them, and no new particles have been
formed. The {\it right panels} show the SFE for Clouds 1 and 2 in the
LAF1 simulation (with feedback).}
\label{fig:SFE_local}
\end{figure*}

The SFE is seen to be reduced by a larger
factor ($\sim 10 \times$) in the case of the LA runs, in which the
collapse is less focused and less massive, than in the case of the SA
ones ($\sim 3 \times$), in which the opposite is true.
In addition, in Fig.\ \ref{fig:SFE_local} we show the evolution of the
SFE at the level of the clouds. The {\it left panel} shows the evolution of the
SFE for the Central Cloud in the SA runs. The {\it middle panels} show
the corresponding plots for Cloud 1 and Cloud 2 in the LAF0 run (without
feedback), and the {\it right panels} show the SFEs in the LAF1 run
(with feedback). Again we see a trend for the less massive cloud (Cloud
1) to suffer a greater reduction of its SFE (by a factor of $\sim 20$,
from $\sim 60$-90\% to $\sim 3$-4\%) than the more massive one (Cloud 2,
by a factor of $\sim 10$, from $\sim 70$-80\% to $\sim 7$-8\%). We
discuss the possible causes for this mass-dependent effect of the
feedback in Sec.\ \ref{sec:accr_vs_feed}. 

\begin{figure}
\includegraphics[width=0.49\textwidth]{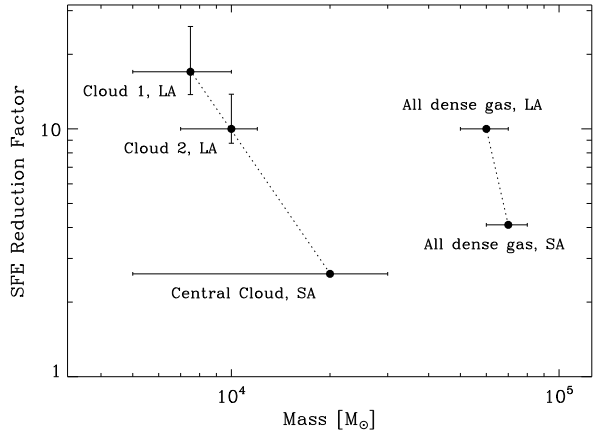}
\caption{Reduction factor of the SFE at 40 Myr upon inclusion of the
feedback in the simulations, both for the three clouds as well as for
the entire mass of dense gas in the simulations. The error bars in the
masses indicate the range of values taken by the systems once their
initial assembly has finished (see Figs.\
\ref{fig:SA_gas_star_mass_evol}-\ref{fig:LA_CL2_gas_star_mass_evol}).
The error bars in the reduction factor for Clouds 1 and 2 denote the
maximum and minimum values of the ratio SFE(w/o
feedback)/SFE(w/feedback), using the SFE data for the three cylinder
sizes of 10, 20 and 30 pc.
}
\label{fig:SFE_vs_mass}
\end{figure}

The factor by which the SFE is reduced upon the inclusion of the
feedback in the simulations is plotted versus the system's mass in Fig.\
\ref{fig:SFE_vs_mass}, both for the full amount of dense gas in the
simulations, and for each of the clouds we have been considering. We see
that two sets of points are clearly defined in this plot, one for the
clouds, and one for the simulations. In both cases, however, the trend
of larger reduction factor at smaller dense gas mass is clearly
observed, although, at the level of simulations, we see that their masses
are not very different. Thus, in this case the different reduction
factors must include a contribution from the larger degree of
fragmentation occurring in the LA simulations due to the larger
amplitude of the initial turbulent fluctuations.
begin{figure}

\begin{figure}
\includegraphics[width=0.7\textwidth]{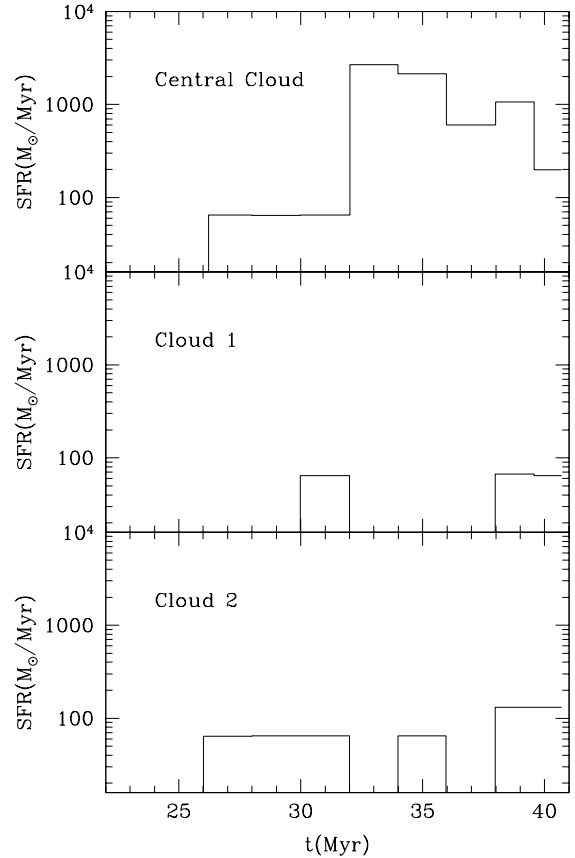}
\caption{Evolution of the stellar particle formation rate, averaged over
2-Myr intervals, of the Central
Cloud ({\it top panel}), Cloud 1 ({\it middle panel}) and Cloud 2 ({\it
bottom panel}). For this plot we only use the 10-pc cylindrical volumes
for Clouds 1 and 2. Because our stellar particles form at a density of
$n = 4 \times 10^6 \pcc$, actual star formation rates should be a factor
of 2-3 lower. Gaps indicate periods over which no new stars are formed.}
\label{fig:SFRnubes}
\end{figure}

\begin{figure}
\includegraphics[width=0.49\textwidth]{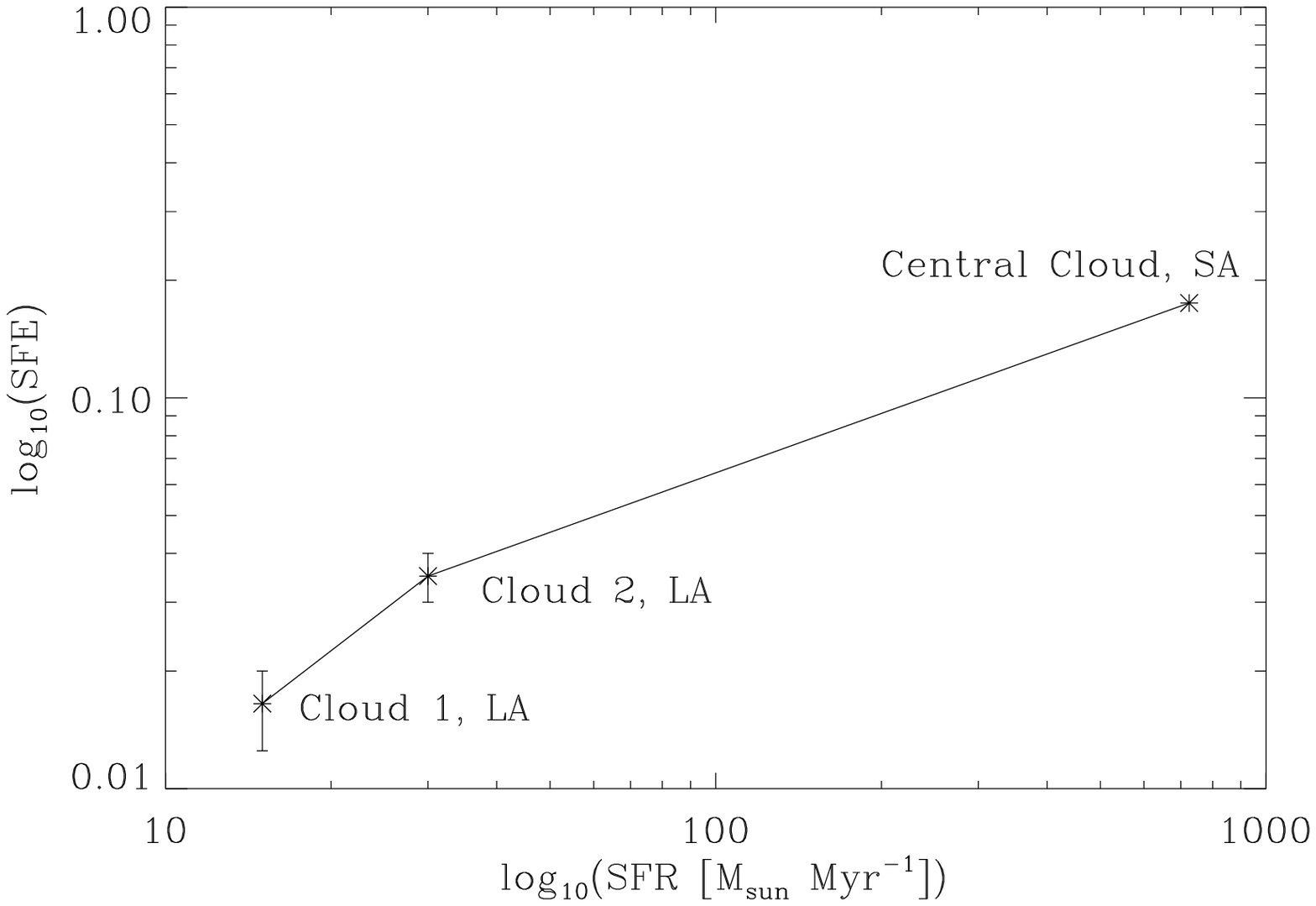}
\caption{SFE {\it versus} stellar-particle formation rate for the three
clouds in the simulations, using the data from Figs.\
\ref{fig:SFE_local} and \ref{fig:SFRnubes}. The values of both the SFR
and the SFE have been multiplied by a factor of 0.5, representative of
the still-lower-than-unity efficiency within our stellar particles.
}
\label{fig:SFE_vs_SFR}
\end{figure}

In order to compare the SFEs in our simulated clouds with those of real
molecular clouds, it is convenient to note that regions of low-mass star
formation, such as Taurus \citep{Goldsmith_etal08} or the Chamaeleon II
dark cloud \citep{Spezzi_etal08} generally have low SFEs ($\sim 1$-5\%),
while cluster forming cores have SFE $\sim 30$-50\% \citep{LL03}. Thus,
we can check whether the SFEs and SFRs of our three clouds follow the
same trend. Figure \ref{fig:SFRnubes} shows the evolution of the SFRs
for our three clouds. From these plots, we find that the average SFR of
the Central Cloud starting from $t=32$ Myr (the time at which a large,
roughly stationary SFR sets in) is $\langle\hbox{SFR}\rangle \sim 1450
~\Msun$ Myr$^{-1}$, while for Clouds 1 and 2, we find
$\langle\hbox{SFR}\rangle \sim 30$ and 60 $\Msun$ Myr$^{-1}$ during
their entire star-forming stages, respectively. From here, and using the
instantaneous SFE at $t=40$ Myr from Fig.\ \ref{fig:SFE_local}, we can
then plot the SFE {\it versus} the SFR. This is shown in Fig.\
\ref{fig:SFE_vs_SFR}, in which both the SFR and the SFE have been
multiplied by an extra factor of 0.5, to respresent the fact that the
efficiency within the stellar particles, which are created at a density
of $n = 4 \times 10^6 \pcc$ in our simulations, is still
smaller than unity. We take 0.5 as a representative value.

We see that the Central Cloud has values of the SFR and the SFE
comparable to those of cluster-forming cores \citep{LL03}. Specifically,
\citet{VS_etal09} estimated an SFR $\ga 250 \Msun$ Myr$^{-1}$ for the
Orion Nebula Cluster (ONC). This calculation used an estimated age spread
of the ONC of $\la 2$ Myr \citep{Hillenbrand97}, and
\citet{Tobin_etal09}'s result of there being 1613 stars in the
ONC. Taking this number as a proxy for the total stellar production of
this region, and a mean stellar mass of $0.3~\Msun$ \citep{HC00}, this
implied a total stellar mass of $\sim 500
\Msun$. On the other hand, \citet{KT07} quote a total stellar mass in
the ONC of $\sim 4600 \Msun$ \citep{HH98} and an age spread of $\sim 3$
Myr \citep{Tan_etal06}, implying an SFR $\sim 1530
\Msun$ Myr$^{-1}$. These values bracket the SFR we measured for our
Central Cloud. 

Concerning the SFE, Fig.\ \ref{fig:SFE_vs_SFR} shows that the Central
Cloud has SFE $\sim$ 10\%, which is comparable to that of the Orion A
cloud \citep{Carpenter00}, in which the OMC-1 clump is contained. Thus,
our Central Cloud may be compared to the Orion A cloud, and its dense
core, discussed in \citet{VS_etal09}, is comparable to the OMC-1 clump.

On the other hand, Clouds 1 and 2 are seen in Fig.\ \ref{fig:SFE_vs_SFR}
to have SFRs comparable to those of low-mass star-forming clouds, $\sim
10$-100 $\Msun$ Myr$^{-1}$, such as Taurus, Cha II or Lupus, in which
the SFE is known to be lower, $\sim 1$-4\% \citep{Spezzi_etal08}. This
compares very well with the SFEs we report for our Clouds 1 and 2 in the
same figure. We conclude that Clouds 1 and 2 are good models of low-mass
star-forming regions, while the Central Cloud is a good model of a
massive SF region, as discussed in \citet{VS_etal09}.

\subsection{The physical processes acting on the clouds}
\label{sec:cloud_conditions} 

\subsubsection{Cloud ``destruction''} \label{sec:destruction} 

Up to the 40-Myr time to which we have evolved our simulations, the
three large clouds (the Central Cloud in run SAF1 and Clouds 1 and 2 in
run LAF1) do not show any instances of the dense gas mass reversing its
increasing trend and beginning to decrease due to the feedback, as can
be seen in Figs.\ \ref{fig:SA_gas_star_mass_evol} through
\ref{fig:LA_CL2_gas_star_mass_evol}. Apparently, the HII region-like
feedback we use is unable to overwhelm the large gravitational potential
well of these clouds and their enveloping ``atomic'' gas reservoirs.

Instead, complete destruction seems to be able to occur in 
small clumps. This can be seen in the animation corresponding to Fig.\
\ref{fig:SA_CC_t265_zoom}, in which various small clumps are seen to be
destroyed by their stellar products. Three particularly conspicuous ones
are, first, the one that forms a stellar particle at the very starting frame
(record 179) of the animation, slightly above and to the right of the
screen's center. Next, another stellar particle appears in the clump
almost at the left border of the frame, about 2/3 of the way from bottom
to top, at record 187. Finally, a third particle appears at record 189,
slightly below and to the right of the screen's center, as the result of
the collision of two clumps. In all of these cases, a single stellar
particle is formed ($\sim 120 \Msun$), and the clump is destroyed. It is
worth noting that actually, the expansion of the HII regions formed
produces new clumps from the material collected around it, but these new
clumps either disperse, or simply do not form new stars. 

Thus, we conclude, similarly to \citet{KMM06}, that small clouds
(``clumps'') are rapidly destroyed, while large clouds may survive for
longer times. However, our clouds exhibit a fundamental difference with
respect to the model considered by those authors, namely that the clouds
are accreting in general. In the next section we now discuss this
feature.

\subsubsection{Accretion {\it vs.} feedback}
\label{sec:accr_vs_feed} 

One crucial feature in all our simulations is that {\it the clouds are
accreting material from the surrounding diffuse medium.} This is
fundamentally different from models in which the clouds are isolated
entities, in rough balance between their self-gravity and the turbulent
pressure, possibly driven by the stellar feedback. The accretion
competes with SF and stellar feedback in regulating the cloud's
mass and coherence, with important consequences. First of all, this
implies that simple observational estimates of the SFE in GMCs based on
measuring the stellar mass and dividing it by the cloud's mass may be
failing to take into account the additional ``raw material'' for SF
contained in the part (or the whole) of the atomic envelope of the
clouds that will eventually be incorporated into the GMC.

Second, the competition between feedback and accretion may explain our
observation from Sec.\ \ref{sec:SFE} that cases with feedback are
characterized by {\it larger} dense gas masses and smaller stellar
masses than their counterparts without feedback. The smaller stellar
mass is not surprising, as the obvious effect of stellar feedback is to
reheat the cold, collapsing star-forming gas, thus reducing the
SFR. However, the larger cold gas mass in the presence of feedback is
indeed surprising, since both gas consumption by SF and the
``ionization'' by stellar feedback act to reduce the dense gas mass. Our
result implies that the rate of dense gas consumption by SF far outweighs 
its rate of destruction by stellar feedback, so that  the net
effect of reducing the SFR is to allow a larger amount of dense gas to be
collected by the accretion. This scenario is supported by Fig.\
\ref{fig:total_mass_evol}, which shows the total (dense gas + stars) in
the clouds in the two sets of simulations. It is seen that the total
cloud mass is nearly the same with and without feedback, suggesting that
{\it the total cloud mass is mainly determined by the accretion, while
the ratio of dense gas to stellar mass seems to be mainly determined by
the feedback.}

These results moreover support the suggestions by \citet{Elm07} and
\citet{VS_etal09} that the majority of the gas in a GMC is not
participating of the SF process at any given time. The latter authors
suggested that this is how local regions of SF may have a very high
specific SFR ($\sim (10~\hbox{Myr})^{-1}$), while
that of their whole parent GMCs may be much smaller \citep[$\sim (300
~\hbox{Myr})^{-1}$; ][]{MW97}. Because it is injected by the newly
formed stars, the stellar feedback acts preferentially on gas that is
about to form stars next. This allows an efficient suppression
of SF, through only a modest fraction of the total available dense gas
being destroyed. That is, if SF is a highly localized process, then it
is possible to achieve significant reductions of the SFR with only a
modest reduction of the total amount of dense gas, by targeting the
destruction precisely at the star-forming gas.

This mechanism may also explain the trend observed in Sec.\
\ref{sec:SFE}, that the SFE is more strongly reduced by the feedback in
cases where the collapsing gas mass is smaller. This may be understood
as a consequence of the fact that stellar feedback is localized, while
the accretion is extended, and more so for greater mass of the
globally gravitationally unstable region that will form the cloud. 
Thus, as we observe in the animation corresponding to
Fig. \ref{fig:SA_CC_t265_zoom}, the low-mass fragments that are
undergoing SF while on their way to the central site of the global
collapse in run SAF1 are easily destroyed by their stellar activity.
Instead, the massive Central Cloud is not destroyed, as it continues to
accrete mass from large distances at a pace that outweighs the local
destruction by stellar feedback. Clouds 1 and 2 in run LAF1, which do
not involve such an extended collapse, are intermediate cases, in which
the clouds are not destroyed by the feedback, but the latter is more
efficient in reducing the SFE.

Thus, we can conclude that
the efficacy of the feedback in destroying the cloud is maximal when its
region of influence is comparable to the spatial extent of the infall,
and is progressively reduced as the latter involves progressively larger
coherence lengths.

\subsection{Physical conditions in the clouds} \label{sec:phys_cond}

Here we consider the global physical properties of the clouds. We
postpone a discussion of the properties of individual clumps within the
clouds for a future study, to be performed at higher resolution. One
such study, including magnetic fields, although not including stellar
feedback, has recently been presented by \citet{Banerjee_etal09}. 

\subsubsection{Density PDFs} \label{sec:PDFs}

It is important to determine the physical conditions in our clouds in
order to assess their degree of realism. One basic diagnostic is the
probability density function (PDF) of the density field. Although it is
well established that in isothermal flows the density PDF takes a
lognormal form \citep{VS94, PNJ97, PV98, NP99, Ostriker_etal99,
Ostriker_etal01, Federrath_etal08}, for non-isothermal flows the
expectation is in general different, with a near-power-law tail
developing at high densities for flows softer than isothermal
\citep[][although see, e.g., Wada \& Norman 2001, 2007]{PV98, Scalo_etal98,
NP99, GVK05, dAB04}, and a bimodal form arising for thermally bistable
flows \citep[][see also the review by \VS\ 2009]{VS_etal00, GVK05, AH05}.

\begin{figure}
\includegraphics[width=0.5\textwidth]{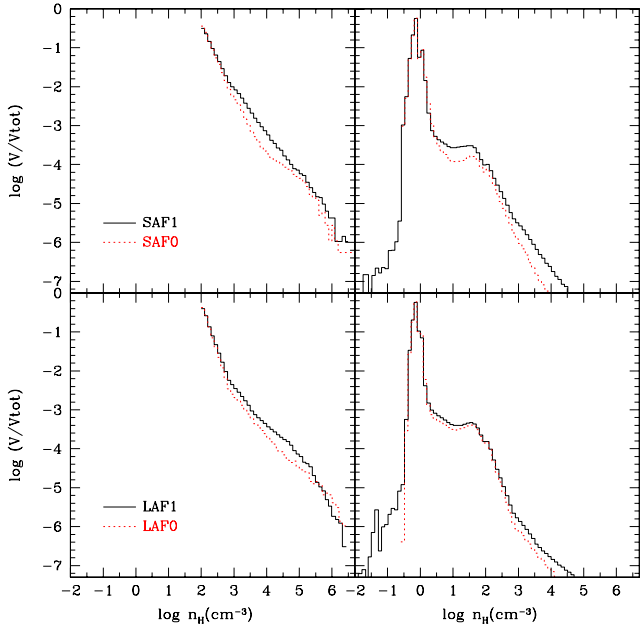}
\caption{Density PDFs for the entire simulation boxes of the SA runs
({\it top panels}) and the LA runs ({\it bottom panels}). The {\it right
panels} show the PDFs for the entire range of densities, while the {\it
left panels} show the PDFs of the dense gas only ($n\ge 100 \pcc$). The
{\it red, dotted} lines refer to the simulations without feedback, while
the {\it black, solid} lines show represent the simulations with feedback.
}
\label{fig:whole_PDFs}
\end{figure}

Figure \ref{fig:whole_PDFs} shows the density PDFs
for the whole simulation box of the SA runs {\it top panels} and of the
LA runs {\it bottom panels}. The {\it right panels} show the entire
density range, while the {\it left panels} show the PDF for the dense
gas ($n\ge 100 \pcc$) only.  
The whole-range PDFs exhibit the bimodality typical of thermally
bistable flows, although the high-density tail is seen to exhibit an
excess over the power-law in both the cases with and without
feedback. This is probably due to the action of self-gravity
\citep{Klessen00, DB05, VS_etal08}. In addition, the cases with feedback
show a slight excess over the cases without it at densities $10^3 \la n
\la 10^5 \pcc$, probably due to the formation of compressed regions by the
expanding ``HII regions''. The high-density PDFs, being just the $n > 100 \pcc$
tail of the whole-range PDFs, only show more detail of the very-high
density gas, up to the star-forming density of $4 \times 10^6 \pcc$, but
with the same shape of the curves as in the whole-range PDFs.

\begin{figure}
\includegraphics[width=0.5\textwidth]{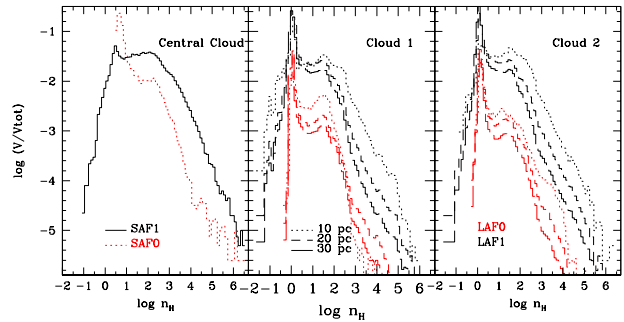}
\caption{Density PDFs for the three main clouds in the simulations: the
Central Cloud in the SA runs ({\it left panel}), and Clouds 1 ({\it
middle panel}) and 2 ({\it right panel}) in the LA runs. The PDFs for the
Central Cloud are computed in a cylinder with length and diameter equal
to 10 pc, while the PDFs for Clouds 1 and 2 are computed in cylinders of
length and diameter equal to 10, 20 and 30 pc. Red lines indicate cases
without feedback, and are displaced downwards by a factor of 10 for
better viewing in the cases of Clouds 1 and 2. Black lines indicate
cases with feedback. 
}
\label{fig:cloud_PDFs}
\end{figure}

\begin{figure}
\includegraphics[width=0.5\textwidth]{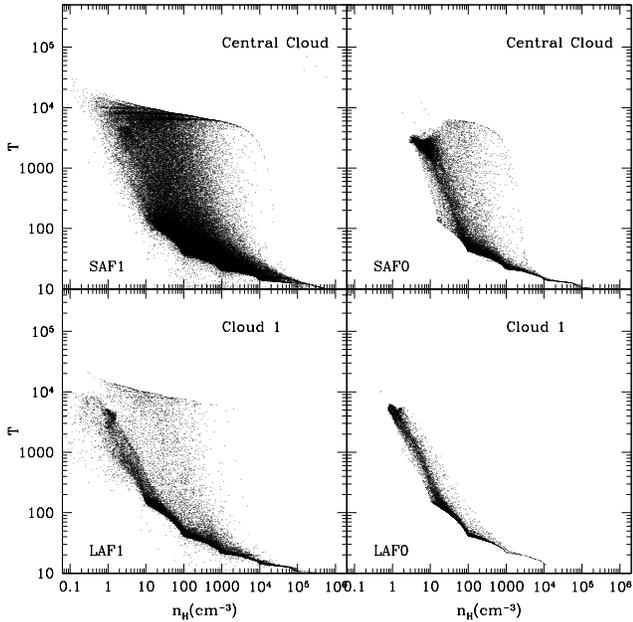}
\caption{Phase-space diagrams of temperature vs.\ density for the
Central Cloud ({\it left panels}), Cloud 1 ({\it middle panels}) and
Cloud 2 ({\it right panels}). The {\it top panels} show the cases without
feedback, and the {\it bottom panels} show the cases with feedback. It
can be seen that, for the majority of the points, the temperature only
varies between $\sim 40$ and 10 K for $100 < n < 10^6 \pcc$. The
exceptions are locations recently heated by new stars.
}
\label{fig:phase_space}
\end{figure}

In order to see the PDFs at the locations of the three main clouds we
have studied (the Central Cloud in the SA runs and Clouds 1 and 2 in the
LA runs), we show in Fig.\ \ref{fig:cloud_PDFs} their corresponding
density PDFs. It is noteworthy that the PDFs again show a roughly
power-law shape at high densities over three to four orders of magnitude
in density, in agreement with the expectation for softer-than-isothermal
flows \citep{PV98, NP99}. This is at odds with results from numerical
simulations of turbulent isothermal gas in closed boxes, but then again
our clouds are {\it not} isothermal. Instead, they are characterized by
an effective polytropic equation of state $P \propto \rho^\gamma_{\rm
eff}$, with $\gamma_{\rm eff}$ varying from 0.8 to nearly unity for $n
\ga 100 \pcc$, as can be seen in Fig.\ \ref{fig:phase_space}. Real
interstellar molecular gas is not expected to be exactly isothermal,
either \citep{Scalo_etal98, SS00, Jappsen_etal05}. Moreover,
our clouds are immersed in a diffuse, warmer medium which, as proposed,
e.g., by \citet{LG03} and \citet{HI06}, infiltrates the clouds to
some degree. Thus, the density PDFs for the volumes containing our
clouds also exhibit the bimodal shape characteristic of thermally
bistable flows, and have low-density tails that extend down to the WNM
regime, contrary to what happens in isothermal simulations.


\subsubsection{Velocity dispersions and virial masses} \label{sec:veldisp}

\begin{figure}
\includegraphics[width=0.7\textwidth]{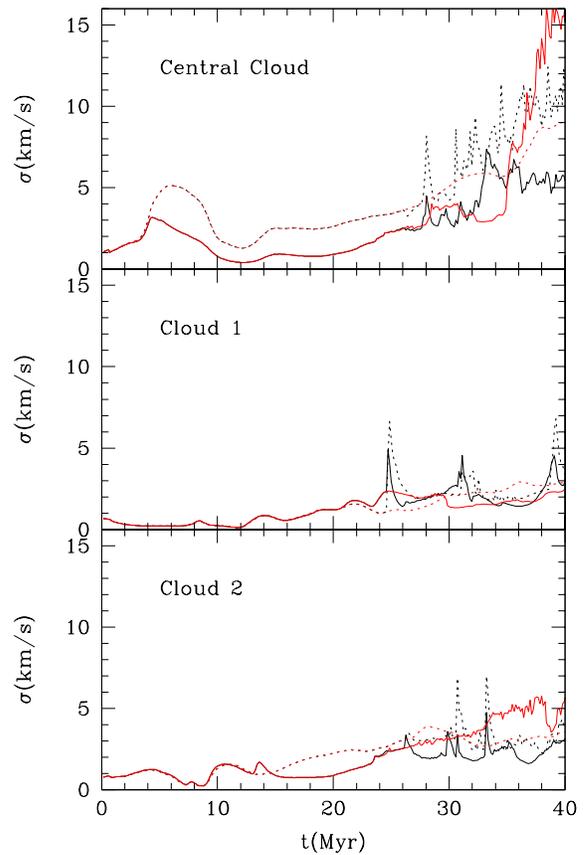}
\caption{Evolution of the velocity dispersion $\sigma$ in the three
clouds. The {\it solid} lines show the density-weighted value, while the
{\it dotted} lines show the volume weighted value. {\it Red} lines
correspond to cases without feedback, and {\it black} lines correspond
to cases with feedback.}
\label{fig:veldisp}
\end{figure}

Finally, we investigate the global velocity dispersion ($\Dv$) in the
Central Cloud and Clouds 1 and 2. This is shown in Fig.\
\ref{fig:veldisp}. The {\it solid} lines show the density-weighted value,
while the {\it dotted} lines show the volume weighted value. {\it Black}
lines denote cases with feedback, and {\it red} lines correspond to
cases without it.

For the Central Cloud and Cloud 2, which are the two most massive ones,
the density-weighted velocity dispersion, which highlights the dense
gas, {\it decreases} upon the inclusion of feedback. Without feedback,
in the Central Cloud, it reaches very high values at the end of the run
($\sim 15 \kms$, corresponding to Mach numbers $\Ms \sim 50$-75), which
in fact are significantly larger than typical values for cloud complexes
of comparable mass \citep[$M \ga 10^4 \Msun$; e.g., ][]{Dame_etal86,
Rathborne_etal09}. Instead, in the run including feedback, $\Dv$ reaches
values $\sim 5$-$6 \kms$ (rms Mach number, $\Ms
\sim 20$), in much better agreement with typically observed values.  On
the other hand, the volume-weighted velocity dispersion, which
highlights the less dense gas, is seen in Fig.\ \ref{fig:veldisp} to
{\it increase} upon the inclusion of feedback. In Clouds 1 and 2, which
are less massive and more scattered than the Central Cloud, $\Dv$ does
not achieve exceedingly large values in any case.

These results can be understood as a consequence of the fact that the
dense gas acquires its largest velocities in the case of free-fall
collapse.  However, the collapse flow is dismantled in its final (fastest)
stages by the stellar feedback, so that it is the high-velocity dense
gas that is preferentially destroyed by the feedback. On the other hand,
this gas becomes warm, diffuse, high-velocity {\it expanding} gas, which
is the one highlighted by the volume weighting. These results again
reinforce the notion that the feedback is mostly applied on the gas that
is closest (but not quite there yet) to forming stars.

Once we have determinations of the velocity dispersion in the cylinders
containing the clouds, it is natural to measure the virial mass
($\Mvir$) of the clouds and compare it with their real mass $M$, in
order to check whether they look ``virialized''.  We compute the
virial mass through the standard formula
\begin{equation}
M_{\rm vir}\equiv 210  \left(\frac{R} {{\rm pc}}\right)
\left(\frac{\Delta v}{{\rm km\ s^{-1}}}\right)^2 M_\odot, 
\label{eq:virial_mass}
\end{equation}
\citep[see, e.g.,][]{Caselli_etal02, Tachihara_etal02, Klessen_etal05}.
For the real cloud mass we consider the sum of the gas (dense + diffuse)
and stellar masses in the volume being considered. It is important to
note that, with this prescription, variations in the value of the ratio
come exclusively from the estimate of the virial mass, because it {\it does}
depend on the weighting used, while the actual cloud mass is
independent of the method of weighting.

\begin{figure}
\includegraphics[width=0.7\textwidth]{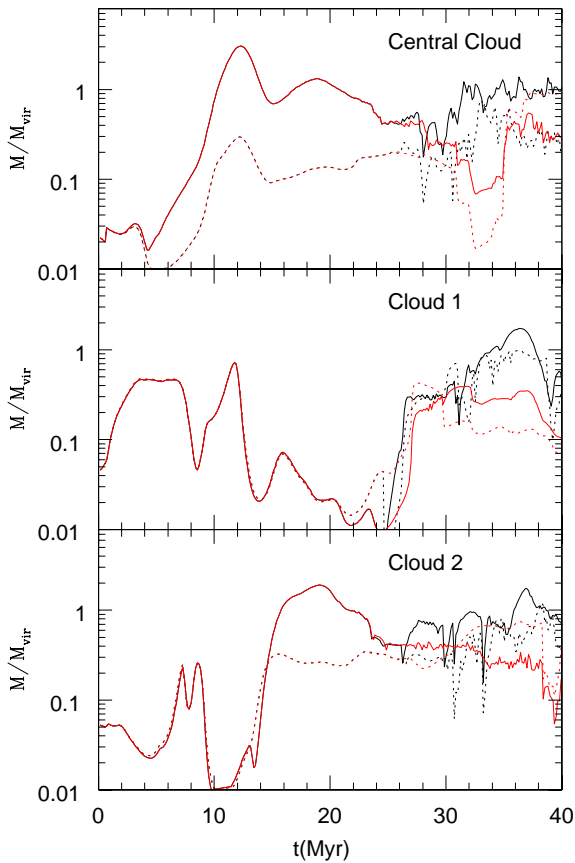}
\caption{Evolution of the ratio of actual to virial mass for the three
clouds. {\it Red} lines correspond to the runs without feedback. {\it
Black} lines correspond to the runs with feedback. {\it Solid} lines
correspond to the density-weighted velocity dispersion, while {\it
dotted} lines correspond to the volume-weighted one. 
}
\label{fig:M_to_Mvir}
\end{figure}

Figure\ \ref{fig:M_to_Mvir} shows the evolution of the ratio $M/\Mvir$
for the three clouds, showing the cases with and without feedback, and
using volume- and density-weighting for the calculation of $\Dv$. In all
cases we observe that, at advanced stages of the evolution, when the
clouds have been fully assembled ($t \ga 25$ Myr), this ratio is closest
to unity when feedback is included and density-weighting is used in the
calculation of $\Dv$. The cases with no feedback and density weighting
are systematically lower than unity by factors $\sim 2$-10. The cases
with volume weighting are more
strongly fluctuating, but it is noteworthy that the ratio often takes
values close to unity anyway.

From these results, we conclude that, in the case with feedback, the
clouds appear to be in a pseudo-virialized state with respect to the
velocity dispersion of the dense gas (highlighted by the density
weighting). In this state, there is an approximate force balance between
self-gravity and feedback driving. However, it differs from true
virialization because of the presence of mass sources (the accretion
from the diffuse environment) and sinks (the consumption by SF and the
destruction by stellar feedback) in the system. These seem to
self-regulate, so as to be capable of maintaining an approximately
constant cloud mass while accretion persists. On the other hand, when
the velocity dispersion of the warm gas is taken into account (still in
the case with feedback), the clouds appear to be slightly sub-virial,
suggesting that the newly formed warm gas in the HII regions is capable
of escaping the cloud.

In the cases without feedback, the clouds also appear to be sub-virial
even with respect to the density-weighted virial mass estimate. In this
case this appears to be a consequence of the dense gas being in
free-fall collapse, and into a potential well produced not only by its
own self-gravity, but also by the previously-formed stars that have
fallen there too \citep[see ][]{VS_etal09}.

\section{Caveats and limitations} \label{sec:caveats}

Of course, our simulations are not free of caveats and limitations. Two
that stand out are the resolution and the nature of the feedback that we
have included. Concerning the resolution, in this paper we have used a
relatively limited one, in order to speed up the simulations, at the
expense of not fulfilling the Jeans criterion proposed by
\citet{Truelove_etal97}. However, as explained in Sec.\
\ref{sec:refinement}, we do not consider this a problem for our study,
since we have restricted it to the global properties of the clouds,
rather than their detailed structure, and we have avoided addressing
issues related with the fragmentation of the final stages of collapse in
the clouds, such as the mass distribution of the cores and stellar
products. We plan to perform a higher-resolution study in a forthcoming
paper, in which we can address these issues.

The second limitation is the nature of the stellar feedback that we have
considered, which in this study has been restricted to local heating
representing the ionizing radiation from massive stars, similarly to the
approach used by, e.g., \citet{YBT82, VPP95, PVP95}, neglecting other
sources such as outflows from stars of all masses \citep[e.g.,][]{NS80,
LN06, Matzner07, NL07, Carroll_etal09, Wang_etal09}. We do this in part
for the numerical simplicity of the approach, and in part because here
we have been mainly interested in the SFE and and cloud evolution at the
scale of GMCs, for which the expansion of HII regions is likely the main
driver \citep{Matzner02}. However, the neglect of bipolar outflows may
introduce a non-negligible bias in our finding that smaller-scale clumps
are destroyed by the feedback, and this should be confirmed by future
simulations that can better resolve these objects and include bipolar
outflows.

\section{Summary and conclusions} \label{sec:conclusions}

In this paper we have presented a numerical investigation of the
evolution of dense (``molecular'') clouds, starting from their formation
by transonic compressions in the warm neutral medium, followed by a
phase transition to the cold neutral medium, the onset of gravitational
collapse, star formation, and, finally, energy feedback from
massive-star ionizing radiation and the formation of expanding HII
regions.

A crucial difference between the results from our simulations and other
models and theories of the self-regulation of star formation and its
efficiency \citep[e.g.,][] {Whit79, Elm83, FST94, MT03, KM05} is that
the clouds in our simulations are in general accreting material at high
rates from the surrounding diffuse medium, rather than having a fixed
mass. This implies that {\it the material making up a cloud is
constantly being renewed over time} because, on the one hand it is
accreting fresh gas, and on the other it is losing mass to SF and
stellar ionization. The star-forming regions can be considered to be not
objects, but rather the {\it loci} where the gas is just ``passing
through'', from the diffuse-gas state to the ``star'' state, similarly
to the nature of a flame, which is the locus of the gas undergoing
combustion in a candle, with fresh air entering at its base, burning
while it transits through the flame, and the exhaust gases leaving at
the top.

Witin this framework, we have found that the SFE in the clouds is
readily decreased by feedback to levels consistent with observational
determinations {\it at all times during the clouds' evolution} up to the
maximum integration time of 40 Myr that we have considered. This is a
significant improvement over our previous non-magnetic studies of the
SFE in the context of cloud evolution without feedback, in which the SFE
at late times is often found to be excessive \citep[e.g., ][]{VS_etal07,
Rosas_etal10}. However, we have found that the reduction factor upon the
inclusion of feedback is an inverse function of the dense gas mass in
the system, and of the degree of coherence of the global collapse, as
illustrated in Fig.\ \ref{fig:SFE_vs_mass}. This result is
complemented by the additional observation that low-mass clumps are
fully destroyed by the feedback.  Together they imply that at some
point, below a certain critical collapsing mass, the appearance of a
stationary state may not occur anymore. This is in qualitative agreement
with the results of \citet{KMM06}, although their model does not include
mass accretion onto the cloud. We plan to present one such model in a
future paper.

In general, our results may provide an explanation for the observational
fact that regions of massive-star formation, which are themselves more
massive than regions of low mass-star formation, appear to have higher
SFEs \citep{LL03} than regions forming low- and intermediate-mass stars
\citep{Evans_etal09}. Essentially, our results imply that {\it the more
massive the region, the less effective the feedback is in reducing the
SFE.} Specifically, these results indicate that the role of the feedback
is not the same in all clouds, but rather depends on the initial
conditions of the large-scale collapse that produces each individual
region. Regions in which the amount of mass involved in the collapse
overwhelms the destructive action of the feedback may reach a stationary
state that appears virialized, but in which there is actually a
countinuous processing of material, since on the one hand the cloud is
accreting mass from the outer infalling material, and on the other hand
it is losing mass through consumption by SF and evaporation by the
feedback. In this case, termination of the SF episode probably requires
the termination of the gas reservoir involved in the global collapse,
and is independent of the feedback. However, this need not be in
contradiction with the low observed global efficiency of SF, if only a
small fraction of a GMC's mass is involved in the collapse at any moment
in time, with the rest being either supported (by, e.g., the magnetic
field), or in the process of dispersal \citep{Elm07}. Alternatively, the
SF episode may be terminated upon the initiation of supernova events,
which we have not included in the present study.

An additional consequence of the mechanisms described in this paper is
that the most massive clouds appear to actually contain more mass when
feedback is included than when it is not. This suggests that, first,
unimpeded SF is more efficient at removing mass from the dense gas phase
to deposit into stars, than the evaporation of dense gas by the stellar
feedback, so that, when the latter inhibits SF, the accretion onto the
cloud accumulates larger amounts of dense gas than otherwise. Second,
this indicates that the deposition of the feedback energy into the
clouds is accurately targeted to gas that is already on the verge of
forming stars. If this gas is a minor fraction of the total dense gas in
a cloud \citep[i.e, SF is a highly spatially intermittent phenomenon in
molecular clouds;][]{VS_etal09}, then its destruction is a highly
efficient way to reduce the SFE without destroying a large fraction of
the cloud.

Finally, we have also investigated the density PDF in the volumes
containing the clouds in our simulations, finding that in general they
retain the bimodal shape characteristic of thermally bistable flows. The
PDF of the dense gas exclusively shows no turnover at low densities,
indicating that low densities are most abundant. The high-density tails
of the PDFs have power-law shapes, as expected for
softer-than-isothermal flows, although this last result may be biased by
our usage of a single cooling law appropriate for atomic gas, and
extrapolating it to molecular-gas densities, rather than using a
specific molecular cooling law. We expect to address this shortcoming in
future works.

\section*{Acknowledgments}
We are grateful to A.\ Kravtsov for providing us with the numerical code,
and to N.\ Gnedin for providing us with the useful analysis and graphics
package IFRIT. The calculations were performed on the 8-core server at
CRyA-UNAM, acquired through UNAM-PAPIIT grant IN112806 to P.C.
This work has received partial financial support from CONACYT grants
U47366-F to E.V.-S. and J50402-F to G.C.G., and UNAM-PAPIIT IN106809 to
G.C.G.

\end{document}